# Improved Mosaicking Capabilities with JEM-X

J. Chenevez[1], N. Lund[1], N.J. Westergaard[1], C. Budtz-Jørgensen[1], P. Kretschmar[2], and R. Walter[2]

[1]*Danish Space Research Institute, Juliane Maries Vej 30, 2100 Copenhagen (Denmark), jerome@dsri.dk*
[2]*INTEGRAL Science Data Centre, Chemin d'Ecogia 16, 1290 Versoix (Switzerland)*

**ABSTRACT**

Observations performed by INTEGRAL are typically subdivided in kilo-seconds individual pointings. A mosaic software has been developed at the Danish Space Research Institute to combine sky images from different pointings with significantly improved capabilities compared to the available software. This software takes account of the specifics of the JEM-X instrument, such as the vignetting effects due to its collimator and its mask structure. It calculates the weighted contribution of each pixel from each input sky images to be added in order to maximise the signal/noise ratio of each pixel in the combined sky image. The software produces mosaic'ed maps with the source signal and of the noise in each pixel, and is used as a tool to find new X-rays sources that are too weak to be seen in individual pointings.

## 1. DEFINITIONS

### 1.1 Vignetting

By use of all available knowledge about one JEM-X detector (its misalignment, its dead areas, its resolution and mask support structure) one calculates an illumination function by a given source for each detector pixel (Brandt et al., 2003). Moving off axis the signal detected from the source diminishes because of the vignetting, first due to the collimator alone, later due to both the collimator and the mask.

### 1.2 Effective exposure

Using the source illumination function one derives a generic map of how the source "recovery fraction" varies as function of sky position. The recovery fraction expresses what fraction of the source flux actually is projected back to the sky position by the reconstruction method. This recovered fraction must be multiplied with the number of pixels used to collect the signal and by the effective observation time corrected for dead time and grey filter effects (Brandt et al., 2003). One has thus a quantity expressing how much additional exposure this map will provide to a combined map.

### 1.3 Variance

The skymap pixel values (the quantities measuring the signal intensity; assuming that all contributions are balanced, the sum of all pixel values across the skymap is zero) are obtained by adding statistically independent counts from a number of points in the shadowgram of the coded mask. Assuming Poisson statistics for many photons, the variance on the total number is just this number. One applies a background subtraction method where counts from the non-illuminated detector pixels are subtracted from the skymap pixel value, or subtracted from a locally estimated background. Because there are much more background pixels than signal pixels, the variance coming from the background subtraction is normally less than the variance on the signal part. It is thus possible to calculate the variance associated to each pixel of a skymap coming from a particular science window.

### 1.4 Skymap pixel weighting

A number of potential contributions coming from skymaps from individual science windows are available for the combined skymap. Each pixel in each contributing skymap is characterised by three numbers: a pixel value, a variance on the pixel signal and a pixel exposure. The weighting of each image depends on the variance in the image, which depends strongly on the number and strength of sources in the field of view (FOV). So the weight factors for a given source depend on the activity level of the surrounding sources. In order to complete the normalisation after the mosaic image is formed, it is also necessary to maintain a fourth and a fifth map containing the sum of the weights and the sum of the weights squared.

When adding a new contribution to the combined map, the S/N ratio in each pixel is maximised in a way that it is always those science windows which have the best exposure to a given pixel that decide the weight of that pixel. This method should thus be optimum for weak source detection purposes.

## 2. GEOMETRICAL ASPECTS

To combine individual skymaps from different pointing observations having typically different orientations in the sky implies some geometrical transformations. First from skymaps in celestial coordinates to JEM-X pixel coordinates (where the weighted mosaicking takes place) and then back to equatorial coordinates for the combined image.

The input skymaps, as well as the mosaic maps are stored in FITS files. The transformation from the FITS image to sky coordinates is described in Calabretta and Greisen (2002).

At the reconstruction stage, the skymap pixels are subdivided into 25 subpixels in order to reduce the distortions due to pixels coming from different orientations. Besides, one can easily show that this pixel subdivision is conservative.

## 3. MOSAICKING

### 3.1 Weighting method

The basics of the reconstruction are shown on Fig. 1. Each input skymap corresponding to a given pointing is composed of a map over the pixel values ($p_j$) and a map over the pixel variances ($v_j$). The effective exposure map ($e_j$) is obtained by multiplying the vignetting map by the effective observation time (available via the *EXPOSURE* keyword).

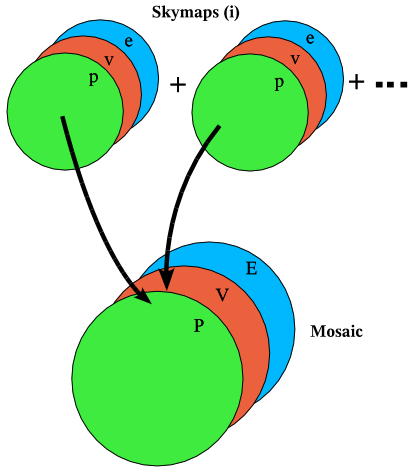

Fig. 1. The basics of the skymaps combining.

In the same way, the combined skymap (mosaic) is composed by the same quantities defined by Eqns. 1-3.

Total pixel value : $P = \sum p_i w_i$ (1)

Total exposure : $E = \sum e_i w_i$ (2)

Total variance : $V = \sum v_i w_i^2$ (3)

Where the weights $w_i$ are obtained in Eqn. 5 by maximising the S/N ratio (Eqn. 4).

$$S/N \sim \frac{E}{\sqrt{V}} = \frac{\sum e_i w_i}{\sqrt{\sum v_i w_i^2}} \quad (4)$$

$$\frac{\partial S/N}{\partial w_j} = 0 \iff w_j = \frac{e_j}{v_j} \frac{\sum v_i}{\sum e_i} \quad (5)$$

### 3.2 Algorithm

At DSRI, our mosaicking software based on the above method uses the following algorithm.
1) Compute the pixel weights as in Eqn. 6

$$\hat{w}_j = \frac{e_j}{v_j} \quad (6)$$

and collect $\sum v_i$ & $\sum e_i$.

2) Apply Eqns 7-9:

$$\hat{p} = \sum p_j \hat{w}_j \quad (7)$$

$$\hat{e} = \sum e_j \hat{w}_j \quad (8)$$

$$\hat{v} = \sum v_j \hat{w}_j^2 \quad (9)$$

3) Then the combined skymaps (Eqns 1-3) are obtained by Eqns 10-12.

$$P = \hat{p} \frac{\sum v_i}{\sum e_i} \quad (10)$$

$$E = \hat{e} \frac{\sum v_i}{\sum e_i} \quad (11)$$

$$V = \hat{v}\left(\frac{\sum v_i}{\sum e_i}\right)^2 \qquad (12)$$

It is also possible, for example, to produce a map over the total weights by applying Eqn 13.

$$W = \hat{w}\frac{\sum v_i}{\sum e_i} \quad \text{with } \hat{w} = \sum \hat{w}_j \qquad (13)$$

Using this algorithm it becomes straightforward that the present process does not depend on the order in which the skymaps are input to the mosaic, which is an obvious constraint for such a combining method.

## 4. PRODUCTS

Producing a mosaic skymap, one may ask onself what are the most relevant quantities to look at. A map of the raw pixel values ($P$), having larger signal values in the centre and smaller signal values at the edges, displays the weighted addition of sources intensity from the different science windows. It is thus the best picture to look at to find faint new sources, especially if they are close to the centre of the FOV. On the other hand, the $P/E$ ratio, as showing the source signal in counts/s/cm$^2$, yields a map where the differences in exposure are compensated. It is thus a more realistic picture showing the relative intensities of the sources. Furthermore, the $\sqrt{(V)}/E$ ratio is a measure of the noise in each pixel and gives a picture of the relative noise over the mosaic map.

The present software can produce mosaic maps of:
- the sum of pointing effective observation times,
- the weighted sum of Intensity (Fig. 2),
- the weighted sum of effective Exposure (Fig. 3),
- the weighted sum of Variance (Fig. 4),
- the sum of weights,
- the sum of weights squared,
- the total Intensity divided by the total Exposure (Fig. 5),
- the ratio $\sqrt{(\text{total Variance})}$ / total Exposure,

as well as the list of input science windows sorted according to their respective weights at a given sky position inside the mosaic image. This may be used to select the best observations of a specific source, which may often be quite off-axis due to dithering pattern constraints.

The software may be used with skymaps from various origins, for it reads the necessary information as regards the images dimension and orientation directly from the input FITS files. By use of a parameter file, it is possible to select the radius of the input skymaps, as well as to define the size of the mosaic image (making possible to magnify a given sky position), and to choose what kind of maps to produce. One may also select different image extensions from the input FITS files, for example if flattened fields skymaps have been produced to reduce the disturbance from eventual strong sources present in the FOV.

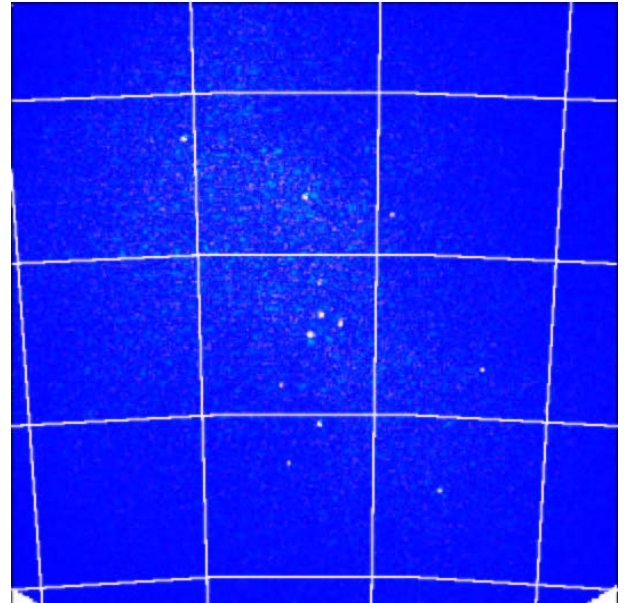

Fig. 2. Raw intensity map of a mosaic of 21 pointings from INTEGRAL revolution 119 showing the Galactic Centre region between 5-10 keV.

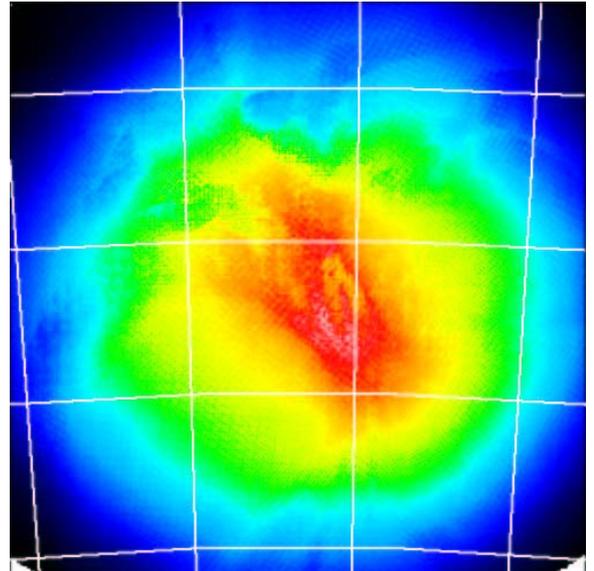

Fig. 3. Same as Fig. 2 showing the effective exposure map. The values decrease from the centre to the edges.

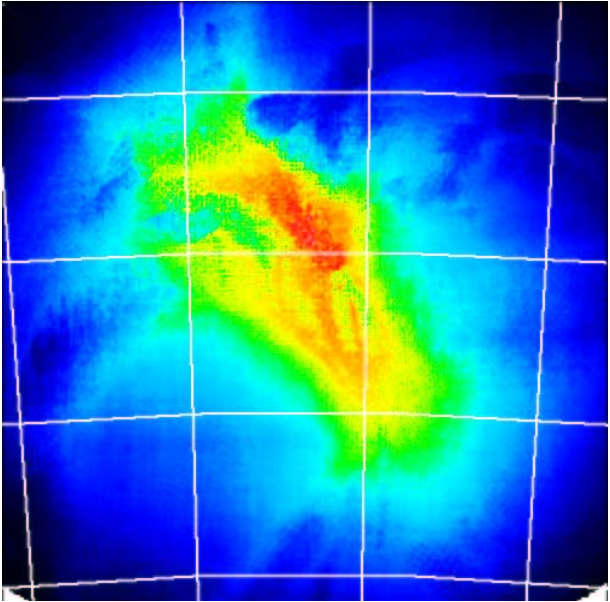
Fig. 4. Same as Fig. 3 for the variance map.

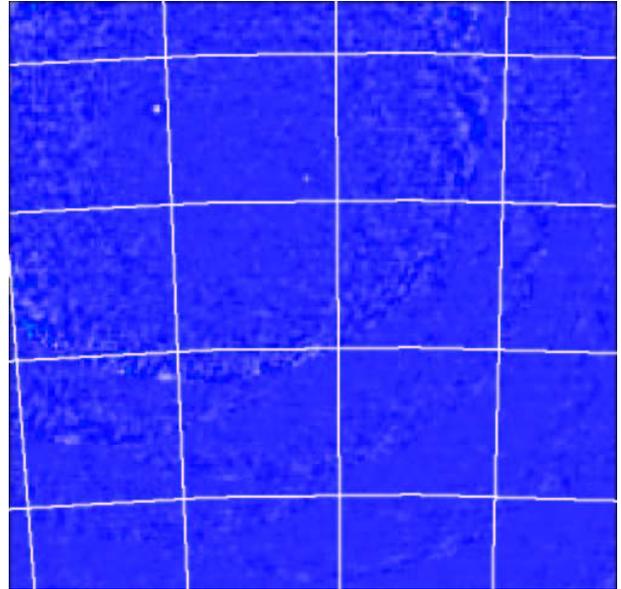
Fig. 6. Intensity map obtained by OSA 3.

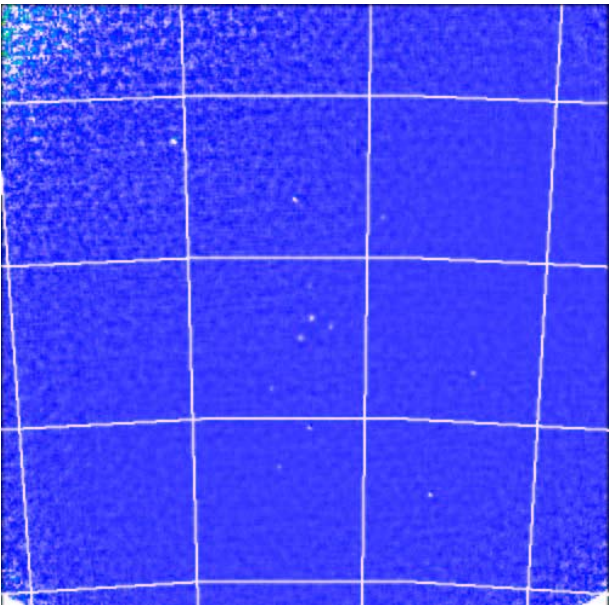
Fig. 5. Map of the intensity divided by the exposure.

As a comparison, Fig. 6 shows the corresponding intensity map obtained by the mosaicking scheme available in the 3rd version of the Offline Scientific Analysis (OSA) software for JEM-X at the INTEGRAL Science Data Centre (ISDC) (Courvoisier et al, 2003). The contrast with Fig. 2 is evident: only a very few X-ray sources are discernible and the background rims of the skymaps contributing to the mosaic image are much more visible, like rings of spurious sources. This demonstrates the benefits of the present approach above a method not based on a weighting combination. Our new method is going to be used in the next generation of the OSA software for JEM-X becoming available from July 2004 at ISDC.

The weighted mosaic images are especially useful to find weak sources. The present software has made possible the detection of the first new source discovered by JEM-X : IGR J06074+2205 (Chenevez et al., 2004). Further developments of the software would include the implementation of an automatic source detection tool.

## 5. REFERENCES

Brandt S. et al., JEM-X inflight performance, A&A 411, L243-L251, 2003.

Calabretta M. R. and Greisen W., Representation of celestial coordinates in FITS, A&A 395, 1077-1122, 2002.

Chenevez et al., ATEL 223, 2004.

Courvoisier T. J.-L., Walter R., Beckmann V., et al., The INTEGRAL ground segment and its science operations centre, A&A 411, L49-L52, 2003.